# CCD Photometry of Dwarf Nova AL Com in Superoutburst

by

Wojciech P y c h and Arkadiusz O l e c h

Warsaw University Observatory, Al. Ujazdowskie 4, 00-478 Warszawa, Poland
e-mail: (pych,olech)@sirius.astrouw.edu.pl

ABSTRACT

We report a CCD optical photometry of a dwarf nova AL Com in superoutburst. Before superhumps occurred the light curve was highly variable with dominant periods about 40 minutes and about 80 minutes for different nights. The period of observed superhumps is 82.5 minutes and seems to be stable. Its first harmonic is also detectable. The 79.3 minutes period is suspected to be a possible signature of orbital motion in the system.

**Key words:** *binaries: close – novae, cataclysmic variables – Stars: individual: AL Com*

## 1. Introduction

Variability of AL Com was first noticed on November 17, 1961 by Rosino (1961) who reported a discovery of a new 14th magnitude object near NGC 4501 (M 88) galaxy. Relative proximity of the star (1950.0 $\alpha = 12^{\rm h}29^{\rm m}54^{\rm s}$, $\delta = 14^\circ 37' 18''$) to NGC 4501 galaxy suggested that it might be a supernova in NGC 4501 or U Geminorum type variable star (Rosino 1961). However the presence of 20th magnitude star in the same position on the POSS plates and large distance from NGC 4501 galaxy caused rejection of supernova hypothesis.

The first light curve of 1961 maximum was reported by Bertola (1964). It indicated that maximum lasted about one month with short-lived deep (3 mag) minimum on December 8. Bertola (1964) also reported the presence of a 14.3 mag maximum in 1892.

Four other outbursts of this star in 1965, 1974, 1975, 1976 are known (Moorhead 1965, Scovil 1975, Howell and Szkody 1987) but there were no short-term modulations during maximum reported.

Spectroscopy and photometry obtained at two first outbursts (Bertola 1964, Moorhead 1965) showed features characteristic for U Geminorum stars in maximum.

CCD photometry of AL Com at quiescence obtained during 5-hour run on April 27, 1987 by Howell and Szkody (1987) showed brightness variations with amplitude 0.4 mag and period about 40 min. This value was confirmed by Szkody *et al.* (1989). Howell and Szkody (1987) and Szkody *et al.* (1989) did not find longer time-scale modulations and suggested that AL Com is likely to be a DQ Her star with white dwarf spin period equal to $42 \pm 1$ min or an AM CVn double degenerate system.

Spectrum of AL Com at minimum brightness shows strong hydrogen lines (Mukai *et al.* 1990). This fact and outbursts of AL Com with amplitude as large as 9 mag make questionable its similarity to AM CVn type stars.

Data obtained by Howell and Szkody (1991) showed modulations with the period of 84 min with no 40 min periodicity. They suggested that AL Com may be a magnetic AM Her system and a switch from two active accretion areas to one active area was observed.

Time-series CCD photometry of AL Com at quiescence from three observing seasons collected by Abbott *et al.* (1992) shows modulations with amplitude 0.2–0.3 mag and two periods: 89.6 and 40.8 min. The 40.8 min oscillation is not the first harmonic of the longer periodicity. During the first observing season Abbott *et al.* (1992) detected oscillation with period equal to 20 min but there was no 90 min periodicity in the light curve.

Although Kholopov and Efremov (1976) reported recurrence time of AL Com outbursts equal to 325 days there were no maxima observed between 1976 and 1994. This behavior together with large amplitude of eruptions makes AL Com similar to WZ Sge (extreme SU UMa stars). Abbott *et al.* (1992) also suggested a similarity of AL Com to the eclipsing intermediate polar EX Hya which shows modulations with 98 min (orbital period) and 67 min (white dwarf spin period).

Beasley *et al.* (1994) presented their microwave survey of magnetic cataclysmic variables. AL Com was classified as DQ Her system with 8.4 GHz emission detected.

The last outburst of AL Com occurred on April 5, 1995 with maximum brightness 12.2 mag (Mattei 1995). The first *V* and *I* bands photometry obtained by Patterson (1995), DeYoung (1995) and Pych and Olech (1995) revealed periods of 81.8, 81.2 and 81.5 min respectively. In this paper we report time-series CCD photometry of AL Com collected on 11 nights during 1995 superoutburst.

## 2. Observations

We observed nova AL Com at the Ostrowik Station of the Warsaw University Observatory on 11 nights from Apr 9/10, 1995 to May 01/02, 1995. A journal of observations is presented in Table 1. The 60-cm Cassegrain telescope with Tektronix TK512CB CCD camera and Cousins filters *V* and *I* were used for the observations (Udalski and Pych 1992). Relative magnitudes $\Delta V$ and $\Delta I$ were obtained as the difference variable minus comparison star. As the comparison star we used 10.5 mag star placed $4'.7$ west of AL Com.

Table 1

Journal of observations of AL Com

| UT Date 1995 | Start (HJD – 2449000.0) | Length of run [hours] | Filter | Integration time [seconds] |
|---|---|---|---|---|
| Apr 09 | 817.531278 | 1.84 | *I* | 60 |
| Apr 10 | 818.309057 | 3.84 | *V* | 60, 90 |
| Apr 11 | 819.289421 | 4.88 | *V* | 90 |
| Apr 20 | 828.309672 | 2.99 | *I* | 90 |
| Apr 21 | 829.311085 | 5.22 | *I* | 90 |
| Apr 22 | 830.393408 | 4.31 | *I* | 90 |
| Apr 23 | 831.387029 | 3.82 | *I* | 90 |
| Apr 24 | 832.304878 | 4.52 | *I* | 90 |
| Apr 25 | 833.413375 | 3.32 | *I* | 90 |
| Apr 30 | 838.386819 | 2.11 | *I* | 120, 180 |
| May 01 | 839.451618 | 1.92 | *I* | 120, 180 |

### *2.1. Data Reduction and Photometry*

The data reductions have been performed with a standard software package available in the Warsaw University Observatory, based on IRAF package. The profile photometry was derived with DAOphot II program. Relative magnitudes of AL Com were obtained using star marked as C1 in the finding chart displayed in Fig. 1. Stars marked as C2, C3, C4 and C5 were used as additional check stars. Results of the photometry were further analyzed with CLEAN algorithm (Roberts *et al.* 1987), Scargle algorithm (Scargle 1982) and using Analysis of Variance Method (Schwarzenberg-Czerny 1989) for coherent periodicities in the light curves.

The mean errors of our measurements were about 0.006 mag for the first night, 0.01 mag for the two *V* band nights, 0.01–0.02 mag for six consecutive nights from April 20th to 26th and 0.03–0.05 mag for two last nights.

## 3. Results

### *3.1. Long Term Behavior After the April 1995 Outburst*

Fig. 2 presents photometric behavior of AL Com observed from Apr 09/10, 1995 to May 01/02, 1995 in filter *I*. General decrease of brightness was about 0.07 mag/day during nights from Apr 20/21 to Apr 25/26, 1995. On nights Apr 30/May 01 and May 01/02 the star was fading faster than 0.8 mag/day. It is clearly visible that outburst lasted at least one month.

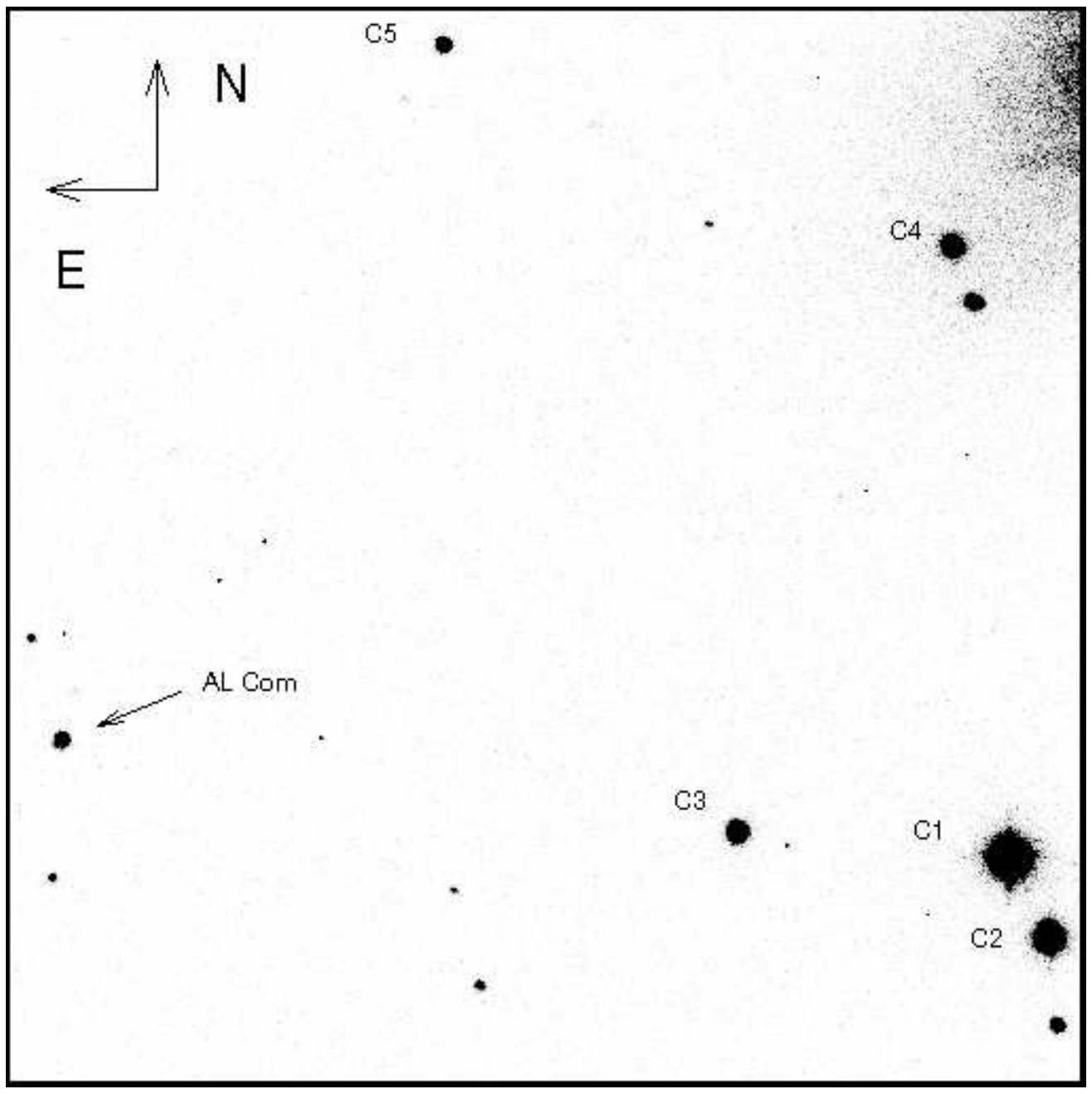

Fig. 1. Finding chart for AL Com observations (6.5 × 6.5 arcmin).

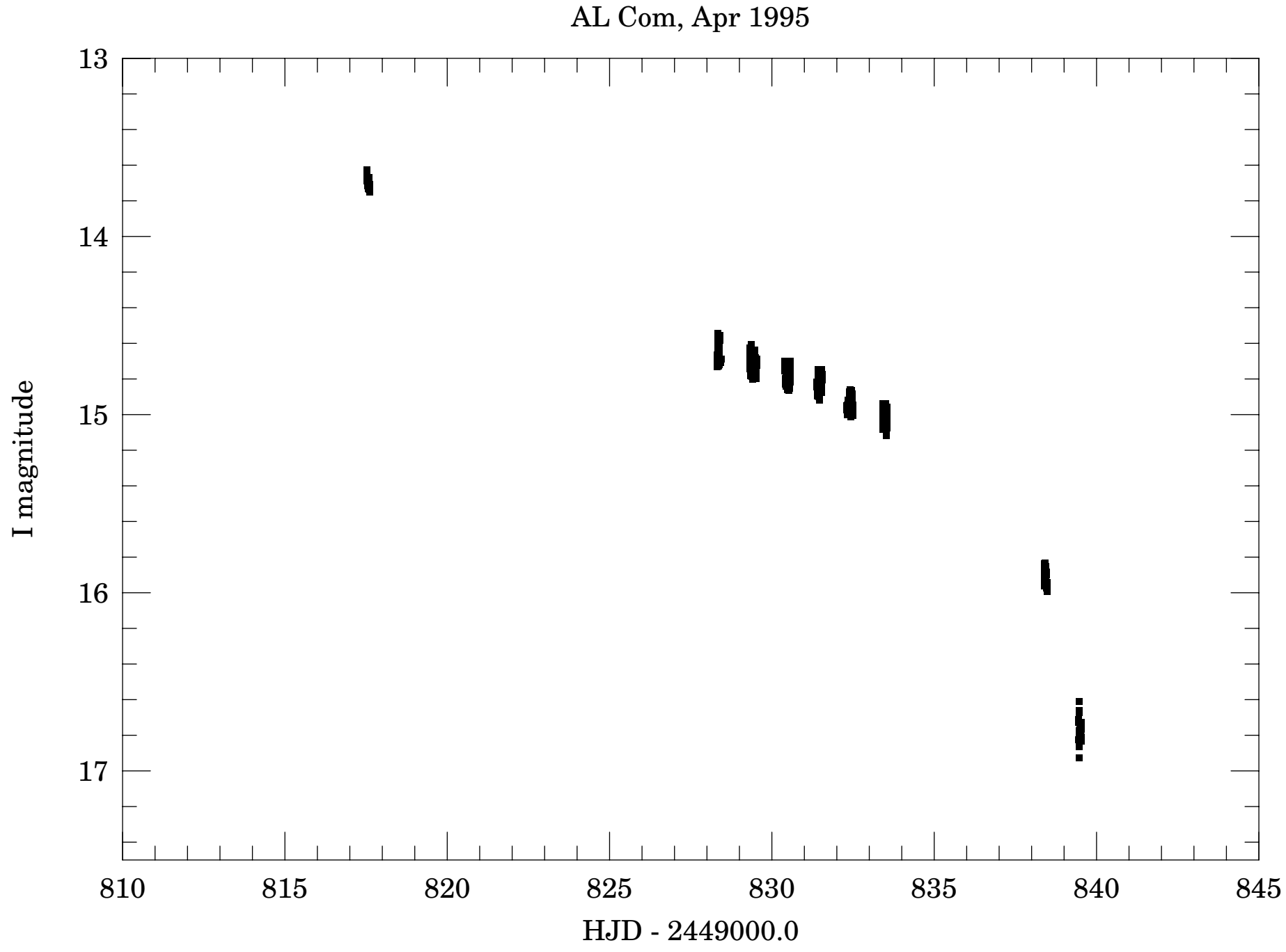

Fig. 2. Photometric behavior of AL Com from Apr 09/10, 1995 to May 01/02, 1995, filter *I*.

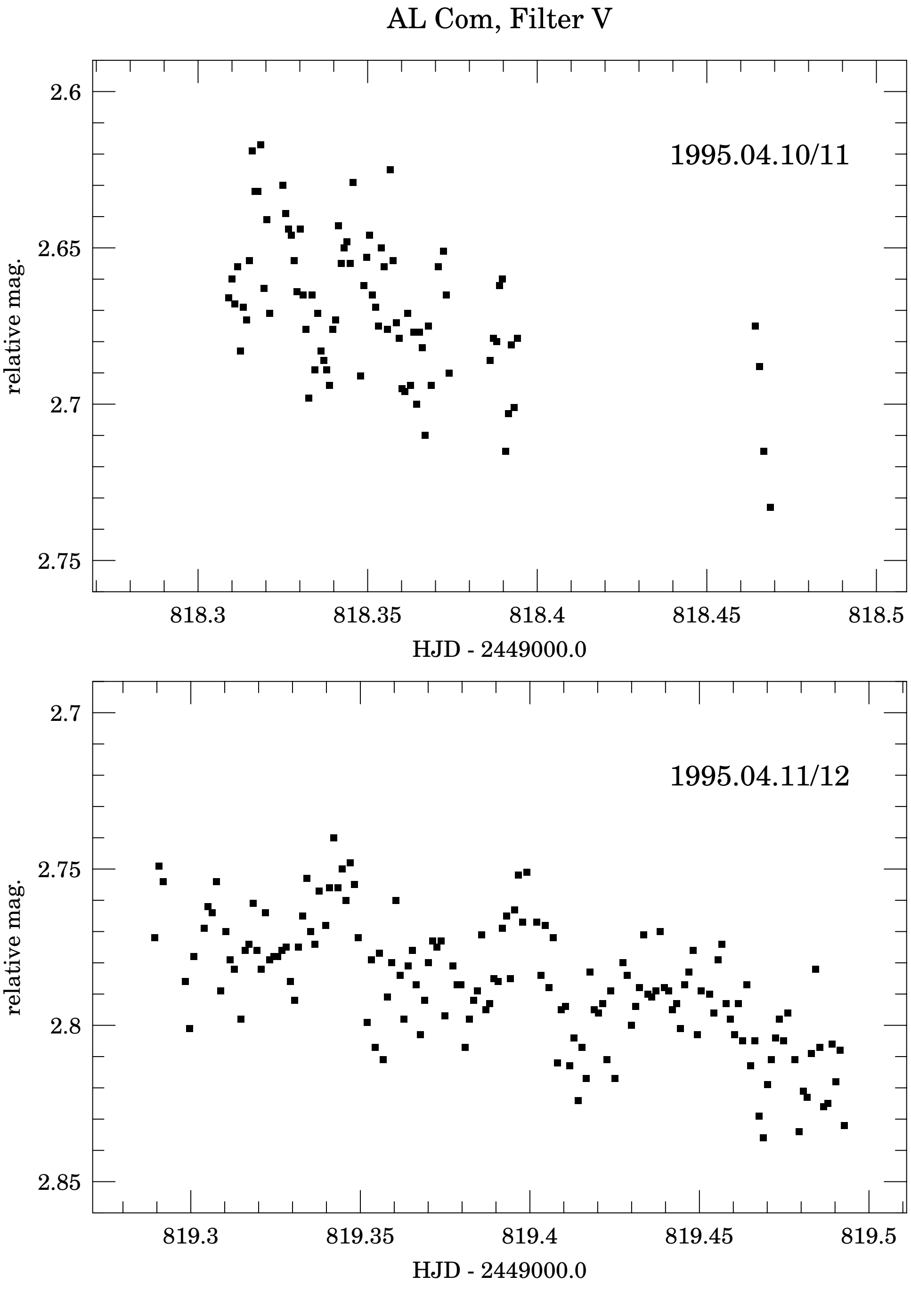

Fig. 3. Light curves of AL Com from Apr 10/11, 1995 and Apr 11/12, 1995 respectively; filter *V*.

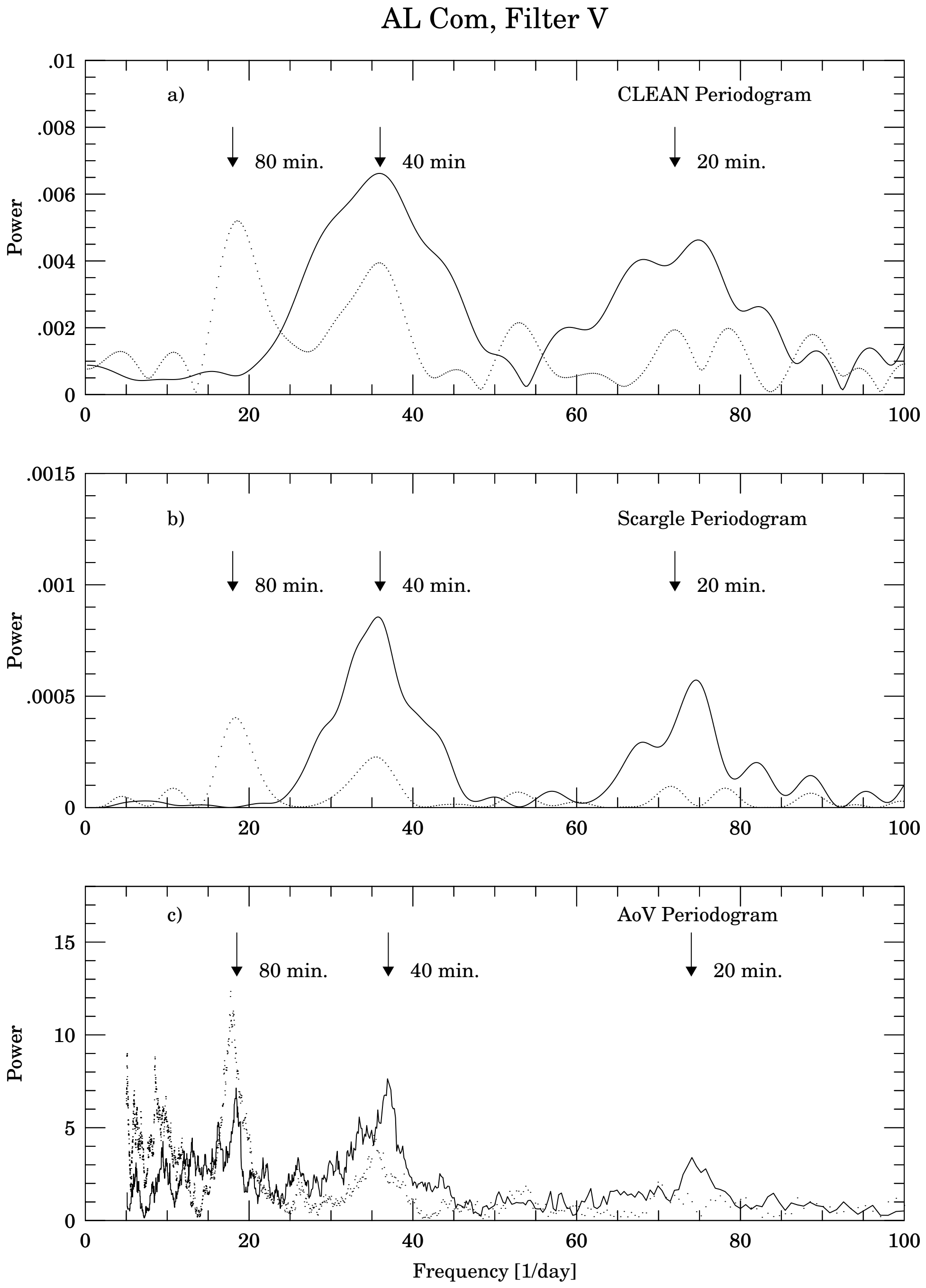

Fig. 4. Periodograms for observations from Apr 10/11, 1995 – solid line and from Apr 11/12, 1995 – dotted line; a) CLEAN spectrum, b) Scargle spectrum, c) Analysis of Variance spectrum.

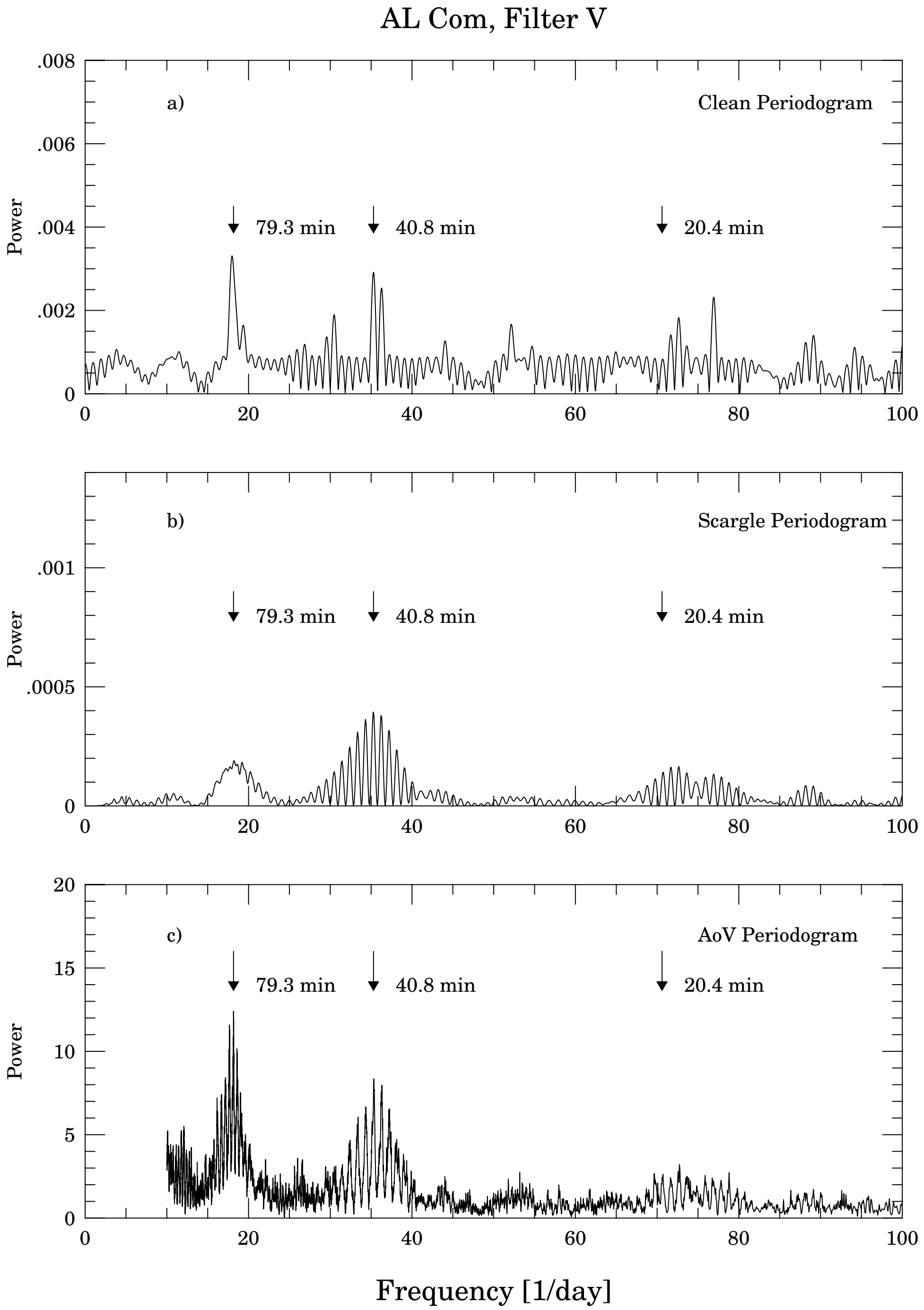

Fig. 5. Periodograms for AL Com, Apr 10/11, 1995 – Apr 11/12, 1995; a) CLEAN spectrum, b) Scargle spectrum, c) AoV spectrum.

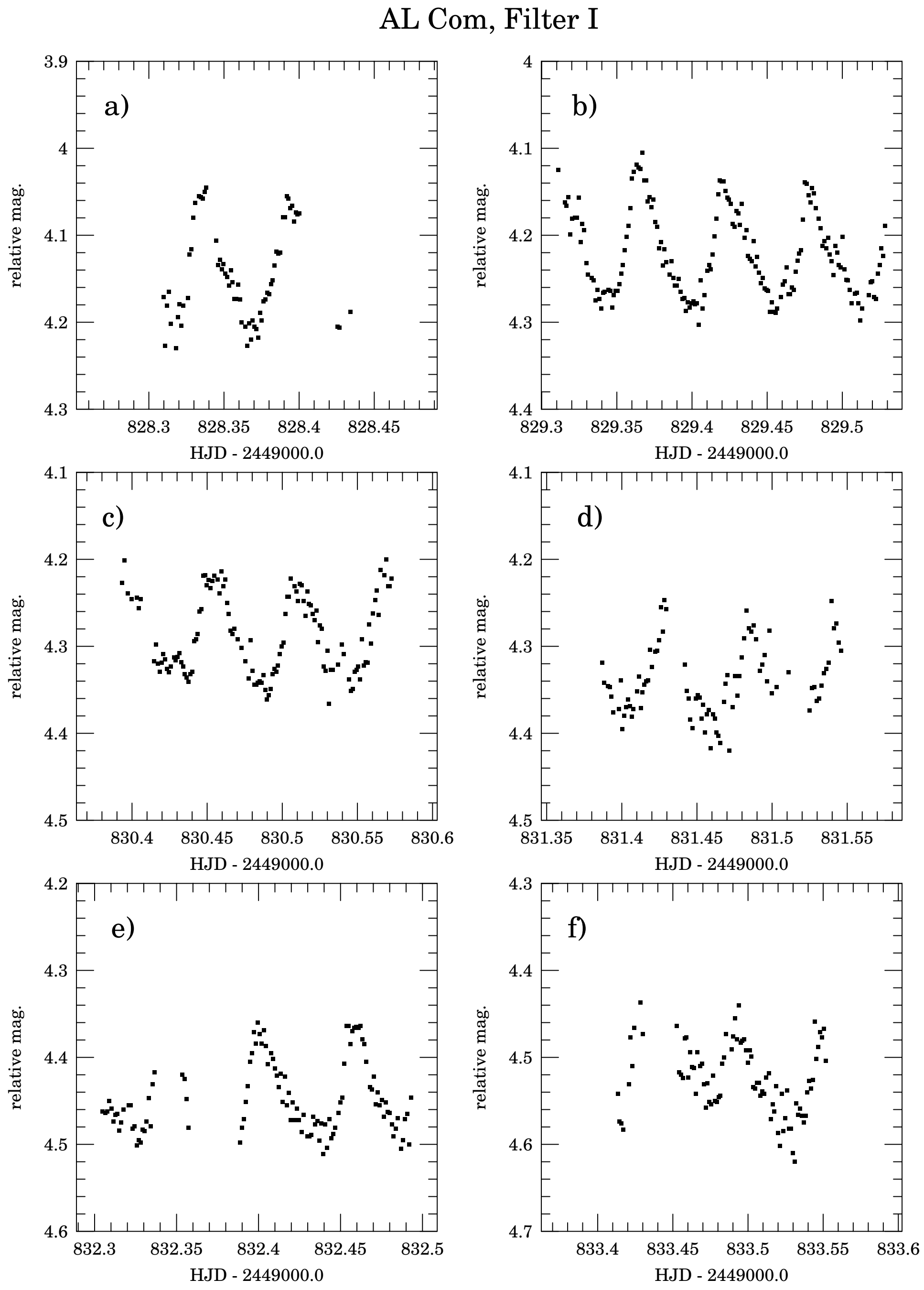

Fig. 6. Light curves of AL Com in outburst. Observations from: a) Apr 20/21, 1995, b) Apr 21/22, 1995, c) Apr 22/23, 1995, d) Apr 23/24, 1995, e) Apr 24/25, 1995, f) Apr 25/26, 1995.

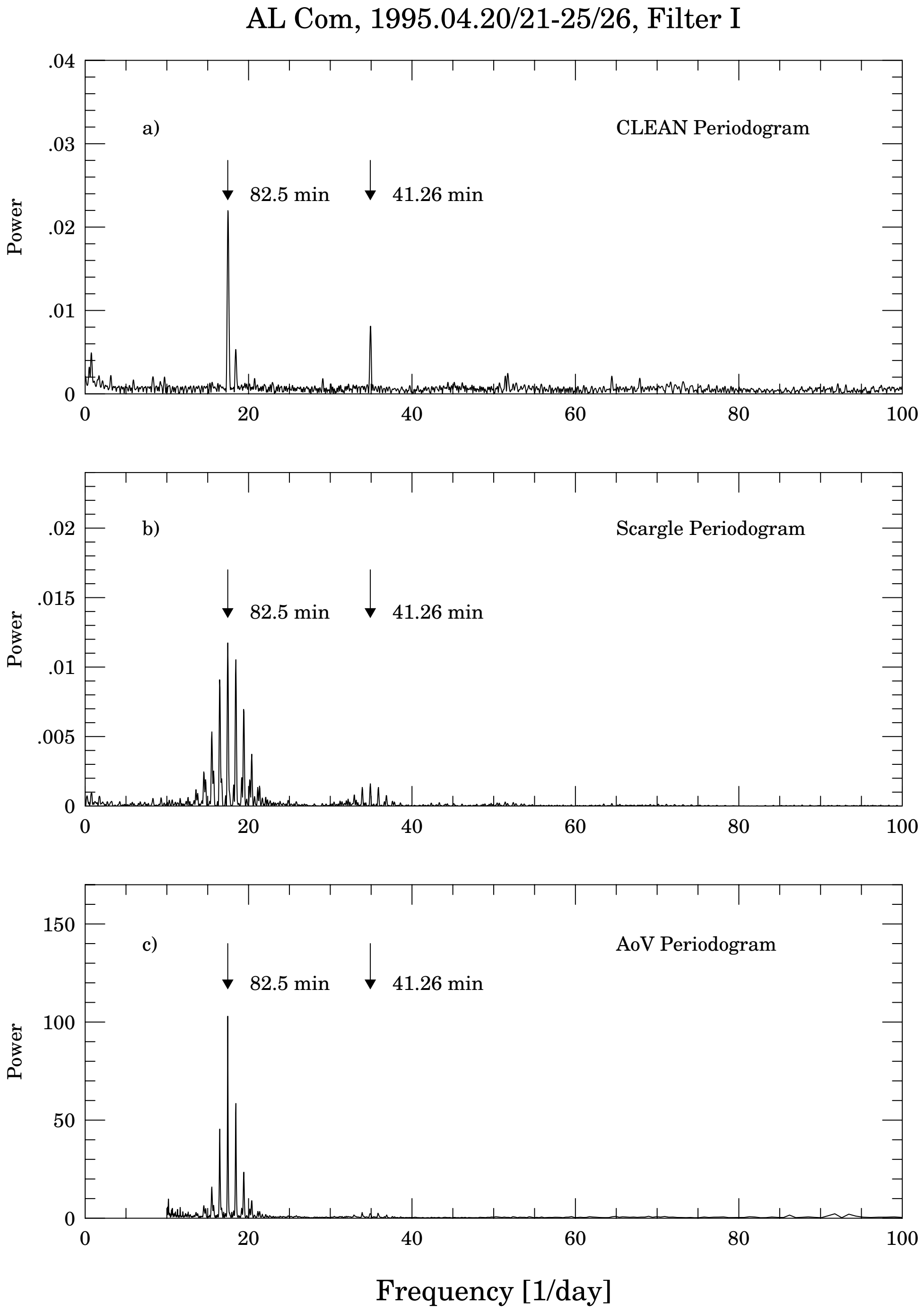

Fig. 7. Periodograms for time-series photometry obtained on nights from Apr 20/21 1995 to Apr 25/26 1995; a) CLEAN spectrum, b) Scargle spectrum, c) AoV spectrum.

### *3.2. Short-term Light Variations*

Pulses observed on Apr 10/11 1995 (filter *V*) had small amplitude of about 0.08 mag. The next night amplitude was even smaller – about 0.06 mag. Fig. 3 presents light curves for these nights.

The light curves exhibit different shape on each night. The first one (Apr 10/11) has peaks of about the same amplitude every 40 min. The second one has smaller amplitude peaks every 79.5 min with secondary ones practically undetectable in our data.

Results of period analysis based on the data from the nights Apr 10/11, 1995 (solid line) and Apr 11/12, 1995 (dotted line) are displayed in Fig. 4.

Differences in light curves from these two nights result in the difference of the periodograms. Frequency corresponding to period of about 40 min is detected in all periodograms. The 80 min peak is a dominant feature in all three spectra for the second night but only the AoV method detected a weak signal for this period in the data from the first night. There is also a wide feature present in the spectra from the night of Apr 10/11 for frequencies close to the first harmonic of the 40 min period.

Fig. 5 presents periodograms based on combined *V* observations on two nights, Apr 10/11 and Apr 11/12, 1995.

From these periodograms we determine periods of $79.3 \pm 1.5$ min for the low frequency peaks and $40.8 \pm 0.5$ min for higher frequency peak. Again, signatures of the possible first harmonic of the 40 min period (about 20 min) are present in the power spectra.

The above variations were observed before superhumps appeared. Fig. 6 presents light curves of AL Com observed during six consecutive nights from Apr 20/21 to Apr 25/26, 1995 in filter *I*. The amplitude of superhumps was about 0.2 mag with minor variations.

CLEAN, Scargle and AoV periodograms (shown in Fig. 7) computed from this set of data yield period $82.5 \pm 0.2$ min and $41.26 \pm 0.09$ min – its first harmonic. The second peak is only a tiny feature on the AoV spectrum, what supports our interpretation that this is the first harmonic and is not a detection of other signal in the light curve.

In order to check the stability of the superhump period (82.5 min) we have analyzed times of the maxima with $O-C$ method. We determined 16 times of maxima from Apr 20 to Apr 26. The best linear fit to these maxima obtained with the Linear Least Squares Method leads to the ephemeris:

$$\begin{array}{rl} \mathrm{HJD_{max}} = 2449828.3345 + & 0.05729\,E \\ \pm\ 0.0020 & \pm 0.00001 \end{array} \quad (1)$$

Table 2 presents $O-C$ data for the maxima of the humps observed from Apr 20/21 to Apr 25/26, 1995 and for period 82.5 min. $O-C$ plot for the times of observed maxima, shown in Fig. 8, indicates no significant period change.

Table 2

$O-C$ for times of superhumps maxima

| HJD –2449000 | Cycle Number | $O-C$ [cycles] |
|---|---|---|
| 828.3345 | 0 | 0 |
| 828.3931 | 1 | 0.0229 |
| 829.3636 | 18 | –0.0370 |
| 829.4200 | 19 | –0.0525 |
| 829.4778 | 20 | –0.0436 |
| 830.4536 | 37 | –0.0110 |
| 830.5084 | 38 | –0.0545 |
| 830.5677 | 39 | –0.0194 |
| 831.4293 | 54 | 0.0199 |
| 831.4850 | 55 | –0.0078 |
| 831.5422 | 56 | –0.0094 |
| 832.3995 | 71 | –0.0452 |
| 832.4580 | 72 | –0.0241 |
| 833.4360 | 89 | 0.0470 |
| 833.4915 | 90 | 0.0157 |
| 833.5489 | 91 | 0.0176 |

The data from nights of Apr 30/May 1 and May 1/2, when the star rapidly faded, became too noisy to be period analyzed, but it seems to be sure that there were no periodic light oscillations larger than 0.1 mag.

## 4. Discussion

Long period between outbursts and shape of the light curve of AL Com during the superoutburst strongly suggest that the star is a member of SU UMa class of cataclysmic variables. Our observations alone would be a good evidence for a SU UMa type variable.

Superhump period derived from our observations is 82.5 min. Two peaks corresponding to 82.5 and 41.25 min periods are present in our periodograms for Apr 20 – Apr 26 (filter *I*) calculated using Scargle and CLEAN algorithms. The peak at 41.25 min period on the AoV periodogram is only a tiny feature. In our opinion the shorter period is just the first harmonic of the basic one.

According to the general relation between superhump period and orbital period for SU UMa stars, the orbital period should be slightly shorter than the period of superhumps (Molnar and Kobulnicky 1992). Using the linear expression relating superhump period with the orbital one (Howell and Hurst 1994) we determine the orbital period to be about 79.5 min.

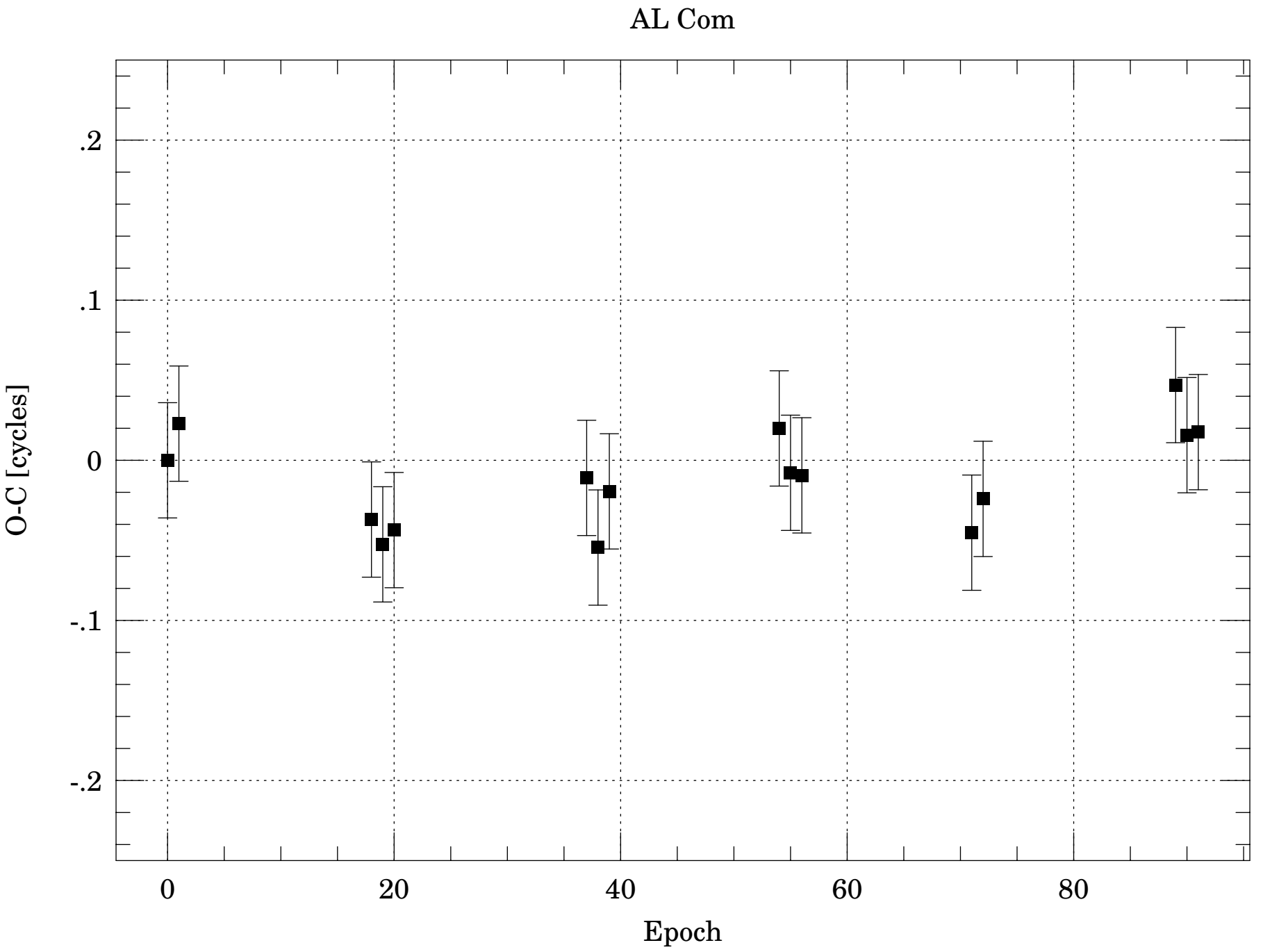

Fig. 8. $O - C$ plot for AL Com maxima observed from Apr 20/21 1995 to Apr 25/26 1995.

The strongest peak in the AoV periodogram for the Apr 10/11 and 11/12, 1995 observations (filter *V*) corresponds to the period of $79.3 \pm 1.5$ min. We are not able to study stability of this period. We think that this period might be the orbital one. However, the possibility that it is a period of the white dwarf rotation, cannot be excluded. This period is one of the shortest orbital periods observed in the SU UMa systems and it is close to the theoretical limit of $81 \pm 6$ min for orbital periods of cataclysmic binaries with nondegenerate secondary (Paczyński 1981). Spectroscopic observations show that AL Com is not an AM CVn double degenerate system (Moorhead 1965, Mukai *et al.* 1990). According to the model of Paczyński (1981) systems with the shortest observed periods should have rapid mass transfer from the secondary to the primary. Instabilities in such flow, in our opinion, may cause observed variations of the light curve shape during quiescence. It also seems that our 79.3 min period may be not the shortest period known for SU UMa stars. O'Donoghue *et al.* (1991) reported WX Cet 1989 superoutburst observations and found superhump period near 80 min. As it is known that the orbital period for SU UMa star is usually a few percent shorter than the superhump period, orbital period of WX Cet should be also shorter than 80 min.

The 84 and 89.6 min periods reported in earlier observations at minimum brightness (Howell and Szkody 1991, Abbott *et al.* 1992) should be then a reflection of a different type of variability – probably white dwarf rotation or an eccentric accretion disk rotation. AL Com was often classified as a magnetic CV. In such a case the superhump period might not be related to the orbital period. For instance if the accretion disk is inclined to the plane of orbit then it would undergo retrograde precession and we would observe humps with period a few percent shorter than orbital one (Patterson *et al.* 1993).

## 5. Summary

Our photometry of AL Com obtained during the 1995 superoutburst suggests that Al Com is a member of WZ Sge subclass of SU UMa stars (O'Donoghue *et al.* 1991). The following points support this interpretation:

- large amplitude ($\sim$ 9 mag) and rare superoutbursts which last more than one month
- presence of short dip of more than 2 mag in the light curves of both WZ Sge during its 1978 superoutburst and AL Com during its 1961 superoutburst
- lack of superhumps during first few days after maximum of brightness observed in both cases
- spectral features of AL Com characteristic for U Gem stars

There is still no satisfactory explanation of the 40 min periodicity which was clearly visible in our *V* band data and was reported by many authors (Howell and Szkody 1987, Szkody *et al.* 1989, Abbott *et al.* 1992). These authors interpreted this period as white dwarf spin period and classified AL Com as DQ Her system. DQ Her systems however do not show outbursts characteristic for dwarf novae. Further photometric, spectroscopic and polarimetric observations are needed for final classification of this object.

**Acknowledgements.** We would like to thank Dr A. Udalski and Dr I. Semeniuk for their helpful comments. We are also grateful to U. Majewska and P. Grzywacz for their cooperation in observations.

## REFERENCES

Abbott, T.M.C., Robinson, E.L., Hill, G.J., and Haswell, C.A. 1992, *Astrophys. J.*, **399**, 680.
Beasley, A.J., Bastian, T.S., Ball, L., and Wu, K. 1994, *Astron. J.*, **108**, 2207.
Bertola, F. 1964, *Ann d'Astrophys.*, **27**, 298.
DeYoung, J.E. 1995, *IAU Circ.*, No. 6157.
Howell, S.B., and Hurst, G.M. 1994, *Inf. Bull. Var. Stars*, No. 4043.
Howell, S.B., and Szkody P. 1987, *P.A.S.P.*, **100**, 224.
Howell, S.B., and Szkody P. 1991, *Inf. Bull. Var. Stars*, No. 3653.
Kholopov, P.N., and Efremov, Y.N. 1976, *Variable Stars*, **20**, 277.
Mattei, J.A. 1995, *IAU Circ.*, No. 6155.
Molnar, L.A., and Kobulnicky, H.A. 1992, *Astrophys. J.*, **392**, 678.
Moorhead, J.M. 1965, *P.A.S.P.*, **77**, 468.

Mukai, K. *et al.* 1990, *MNRAS*, **245**, 385.
O'Donoghue, D., Chen, A., Marang, F., Mittaz, J.P.D., Winkler, H., and Warner, B. 1991, *MNRAS*, **250**, 363.
Paczyński, B. 1981, *Acta Astron.*, **31**, 1.
Patterson, J., Thomas G., Skillman, D.R., and Diaz, M. 1993, *Astrophys. J. Suppl. Ser.*, **86**, 235.
Patterson, J. 1995, *IAU Circ.*, No. 6157.
Pych, W., and Olech, A. 1995, *IAU Circ.*, No. 6162.
Roberts, D.H., Lehár, J., and Dreher, J.W. 1987, *Astron. J.*, **93**, 968.
Rosino, L. 1961, *IAU Circ.*, No. 1782.
Scargle, J.D. 1982, *Astrophys. J.*, **263**, 835.
Schwarzenberg-Czerny, A. 1989, *MNRAS*, **241**, 153.
Scovil, C. 1975, *IAU Circ.*, No. 2760.
Szkody, P., Howell, S.B., Mateo, M., and Kreidl, T.J. 1989, *P.A.S.P.*, **101**, 899.
Udalski, A., and Pych, W. 1992, *Acta Astron.*, **42**, 285.